\documentclass[11pt]{article}
\usepackage[utf8]{inputenc}
\usepackage[english]{babel}
\usepackage{sectsty}
\usepackage{graphicx}
\usepackage{amsmath}
\usepackage{amssymb}
\usepackage{comment}
\usepackage{multirow}
\usepackage[numbers]{natbib}
\usepackage{makecell}
\usepackage{tabularx}
\usepackage{subfig}
\usepackage{float}

\graphicspath{{figures/}}
\usepackage[hidelinks]{hyperref} 
\usepackage[capitalise]{cleveref}

\title{\textbf{Bayesian calibration of Arterial Windkessel Model}}

\author{Michail Spitieris$^1$, Ingelin Steinsland$^1$ \& Emma Ingestr{\"o}m$^2$}
\date{
	{\footnotesize$^1$Department of Mathematical Sciences, Norwegian University of Science and Technology (NTNU), Norway\\%
	$^2$Department of Circulation and Medical Imaging, Norwegian University of Science and Technology, Norway\\[2ex]}
}

\makeatletter
\newcommand*\bigcdot{\mathpalette\bigcdot@{.5}}
\newcommand*\bigcdot@[2]{\mathbin{\vcenter{\hbox{\scalebox{#2}{$\m@th#1\bullet$}}}}}
\makeatother

\begin{document}
	\maketitle	
	
	

\section*{Abstract}

This work is motivated by personalized digital twins based on observations and physical models for treatment and prevention of Hypertension. The models commonly used are simplification of the real process and the aim is to make inference about physically interpretable parameters. To account for model discrepancy we propose to set up the estimation problem in a Bayesian calibration framework. This naturally solves the inverse problem accounting for and quantifying the uncertainty in the model formulation, in the parameter estimates and predictions. We focus on the inverse problem, i.e. to estimate the physical parameters given observations. The models we consider are the two and three parameters Windkessel models (WK2 and WK3). These models simulate the blood pressure waveform given the blood inflow and a set of physically interpretable calibration parameters. The third parameter in WK3 function as a tuning parameter. 
The WK2 model offers physical interpretable parameters and therefore we adopt it as a computer model choice in a Bayesian calibration formulation. In a synthetic simulation study, we simulate noisy data from the WK3 model. We estimate the model parameters using conventional methods, i.e. least squares optimization and through the Bayesian calibration framework. It is demonstrated that our formulation can reconstruct the blood pressure waveform of the complex model, but most importantly can learn the parameters according to known mathematical connections between the two models. We also successfully apply this formulation to a real case study, where data was obtained from a pilot randomized controlled trial study. Our approach is successful for both the simulation study and the real cases.


\section{Introduction}
The recent advances in digital technologies have brought an increasing opportunity for personalized medicine. Sensor technology allows for real-time personalized data acquisition while physical simulator models can provide a better understanding of the phenomena under study and assist users for future decisions. These technologies are accommodated in infrastructures called  digital twins  \cite{boschert2016digital}. Many physical simulators are described by (systems of) differential equations which are implemented in computer code and are referred to as computer models. The computer models usually consist of a set of variable inputs which can be measured by field experiments (e.g. sensor data) and a set of  calibration parameters which are initially unknown and are chosen based on calibration to the observed data. 

This paper is motivated by digital twins for the treatment and prevention of hypertension. High blood pressure is a leading modifiable cause of cardiovascular disease, disability and premature death 
\cite{forouzanfar2017global,lim2012comparative,forouzanfar2016global,danaei2014cardiovascular,prospective2002age,rapsomaniki2014blood}.
Computer models providing insight to the underlying hemodynamics may be used to guide personalized treatment, thereby reducing medical side effects, and contributing to more effective treatment strategies of high blood pressure. In this paper, we focus on deterministic models that describe blood pressure in the aorta in terms of blood flow and two key physically interpretable calibration parameters, total vascular resistance and arterial compliance.

Windkessel models \cite{westerhof2009arterial} describe the blood pressure waveform during the cardiac cycle (the time between two consecutive heartbeats). The most basic model, the two parameters windkessel model (WK2), describes this process in terms of vascular resistance and total arterial compliance. An attractive property of the WK2 model is that the two parameters describe important physiological features for hypertension \cite{pettersen2014arterial}. However, it was found that this model does not fit the observed data well especially during systole \cite{segers2008three}, where blood pressure reach its maximum value. To improve the fit to the observations more complex models with an increasing number of parameters  have been proposed in the literature \cite{westerhof1971artificial, stergiopulos1999total, westerhof2009arterial}. More complex models give a better fit to data, but often comes with parameters identifiability problems and/or loss of physical interpretability \cite{segers2008three}. For example, although the three parameters windkessel model (WK3) \cite{westerhof1971artificial} it fits observed data better, it generally overestimates the total arterial compliance \cite{segers2008three}.

Nevertheless, WK models can be useful for estimating parameters with physical meaning (e.g. total arterial compliance) and for simulating the blood pressure waveform under various scenarios. A typical approach for estimating windkessel models calibration parameters is minimizing a loss function, which is the difference between the observed and (simulator) predicted blood pressure given the observed inflow in time \cite{westerhof2009arterial,segers2008three}. Other proposed methods in the literature co-exist and offer points estimates as well, see for example \cite{westerhof2009arterial}. However, in a digital twin setting individual sensor data are usually noisy and physical variability of the process as well, do not agree with a unique set of "best-fitting" calibration parameters. Since windkessel models are just approximations of the real process, one other important source of uncertainty that should be recognized is the model discrepancy. 

Kennedy and O'Hagan (KOH) \cite{kennedy2001bayesian} proposed a Bayesian Calibration framework for dealing with potentially imperfect models and noisy observations. In their formulation they allow the model to have some systematic functional bias, which is modeled by a flexible Gaussian process (GP) prior. For the calibration parameters, KOH suggest incorporating the underlying knowledge in form of prior distributions based on prior knowledge about the processes. The computer model which is usually computationally intensive is modeled by a second Gaussian process model known as emulator (or surrogate) and suggested to infer all unknowns jointly, using Markov chain Monte Carlo (MCMC). Bayarri at al. \cite{bayarri2007framework} proposed a modularized KOH framework where the (GP) emulator hyper-parameters are estimated only from computer model runs. They showed that this can improve identifiability and mixing of the MCMC algorithm. 

In this paper we propose to use the WK2 model which offers interpretability
of the physical parameters. To account for model discrepancy we utilize the modularized KOH framework \cite{bayarri2007framework}. First we consider a synthetic example where the ground truth is a more complex model which has known mathematical connections with the WK2 model. We simulate data from the more complex model, the WK3 model and we add realistic noise for blood pressure measurements. We use the modular approach \cite{bayarri2007framework} to estimate the physical parameters of interest and quantify their uncertainty. Our approach is also demonstrated by a case for real blood pressure and blood flow observations. 

The remaining of the paper is organized as follows. In \Cref{sec:WK} we give a brief introduction to the WK2 and WK3 models including some of their mathematical properties and connections. In \Cref{sec:BC} we review Bayesian calibration method and approaches for dealing with functional output. In \Cref{sec:BC_WK} we introduce our proposed modeling framework, Bayesian Calibration of Windkessel models and inference procedure for blood pressure, as well as the setup of the synthetic and real case study. In \Cref{sec:conclusions} we conclude with a discussion. 

\section{Windkessel Models}
\label{sec:WK}
At every heartbeat, the heart pumps out blood through the aorta. This cause increased pressure in the arteries up to some maximum value (systolic pressure) and is followed by an approximately exponential decay until it reaches a minimum (diastolic pressure). The windkessel two elements (WK2) model \cite{westerhof2009arterial} is given by the following differential equation

\begin{equation}
I(t) = \frac{1}{R}P(t) + C \frac{dP(t)}{dt},
\label{eq:WK2}
\end{equation}

\noindent where $P(t)$ is the pressure and $I(t)$ is the inflow. The compliance, $C$ and resistance, $R$ of the vessels are the unknown parameters and estimated by using the available observed $P(t_i)$ and $I(t_i),$ $i=1,\dots,n$. 

The WK2 model can be viewed as an electric circuit with a resistor (which resists the current) and a capacitor (which stores energy) placed in parallel ($P(t)$ is the voltage that drives the circuit and $I(t)$ is the current). By applying Kirkhoff laws one can derive equation \cref{eq:WK2}. During diastole, the aortic valve is closed and consequently, the blood inflow is 0. This results to an exponential decay of blood pressure, $P(t) = P(t_0) \exp (-\frac{t-t_0}{RC}),$ where $t_0$ is the time point that diastole started.

\begin{figure}[htbp]
	\centering
	\includegraphics[width=0.7\textwidth]{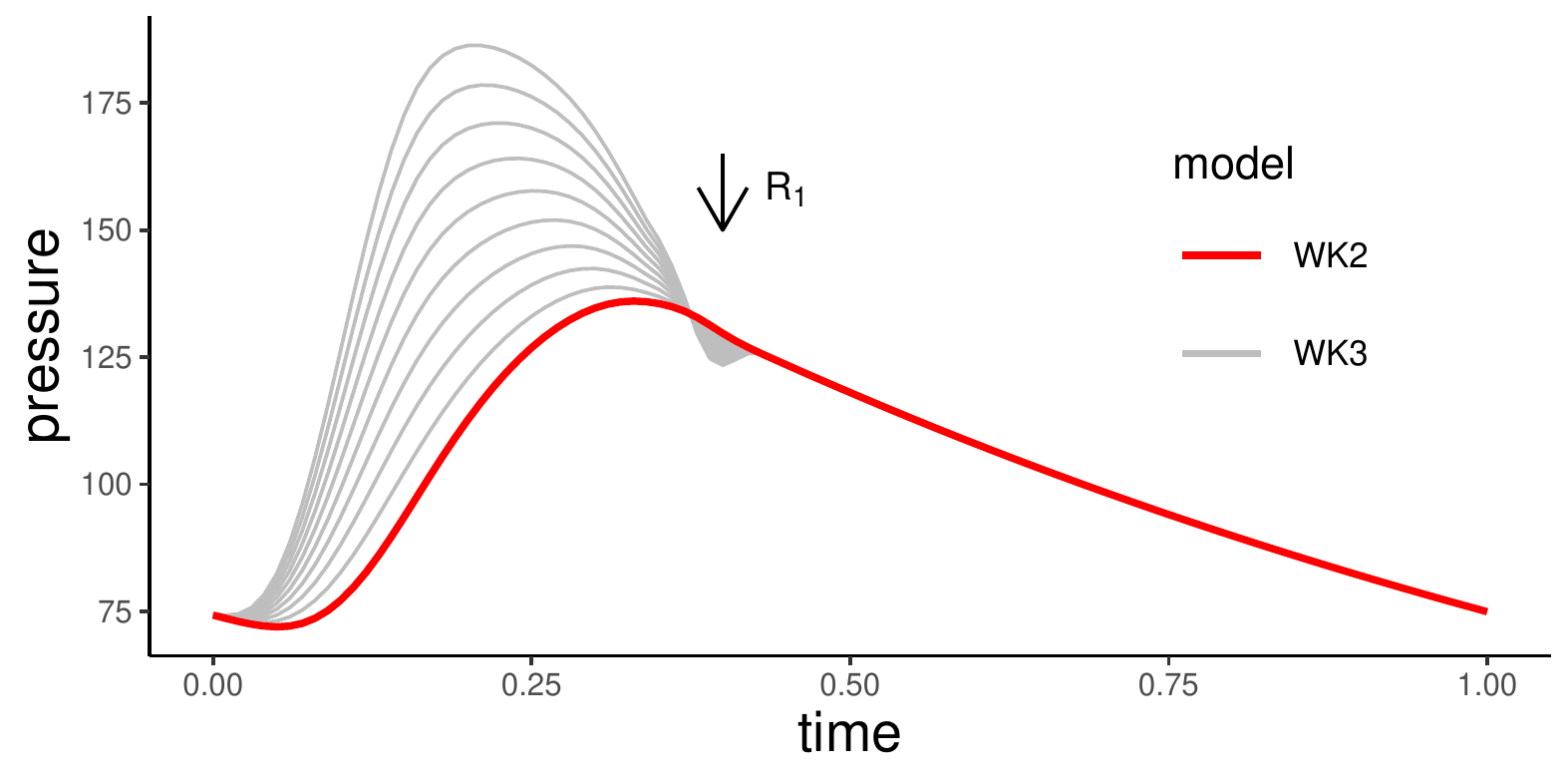}
	\caption{Blood pressure generated from WK2 model (red) and for a range of range of $R_1$ values $[0.01,0.2]$ from WK3 model (grey). The inflow and $C$ values are identical for both models. The amplitude of WK3-generated curve decreases linearly with $R_1,$ while the models become equivalent for $R_1 = 0.$}
	\label{fig:WK2and3}
\end{figure}

Although this model describe the main functions of the aortic pressure, it was found that it can perform poor during systole \cite{westerhof2009arterial}. Wasterhof et al. \cite{westerhof1971artificial} suggested the inclusion of a second resistive element, placed in series with the resistance and compliance of WK2 model. Again by applying Kirkhoff laws we can derive the WK3 equation 
\begin{equation}
\frac{d P(t)}{d t} + \frac{P(t)}{R_2C} = \frac{I(t)}{C} \left (1 + R_1/R_2 \right ) + R_1 \frac{d I(t)}{d t}.
\label{eq:WK3}
\end{equation}
The new element is the characteristic impedance, $R_1$ and can improve the fitting significantly. However it can lead to an overestimation of total arterial compliance \cite{segers2008three}. During diastole ($I(t_d) = 0$)), the decay of blood pressure is equivalent to WK2 and as $R_1 \to 0$ the models become equivalent (see \cref{fig:WK2and3}).

From a modeling perspective $R_1$ controls the amplitude of the blood pressure curve (see \cref{fig:WK2and3}) and its estimated value is usually low ($\approx 0.1$). In this context one can see $R_1$ as a tuning parameter, rather than a parameter with direct physical interpretation. Furthermore an important connection between the two models is that in WK2 model $R$ equals to the ratio of mean pressure over mean flow, while for the WK3 model this ratio is equal to $R_1 + R_2$ \cite{westerhof2009arterial}.

\section{Bayesian calibration}
\label{sec:BC}

Computer models typically consist of variable inputs $\mathbf{x} = (x_1,\dots,x_p)$ which controls the process and can controlled or/and be measured, unknown calibration parameters $\mathbf{u} = (u_1,\dots,u_m)$ and outputs $\mathbf{y} = (y_1,\dots , y_n).$ Often both input and output are continuous in time, and hence are functional \cite{bayarri2007computer}. We typically want to estimate the parameters $\mathbf{u}$ based on measured input and output variables. This referred to as calibrating the computer model. For example, in our case we observe inflow and pressure along with time and we want to estimate the unknown calibration parameters $R$ and $C.$  It is common that computer models do not fit the observed (field) data perfectly, and systematic model bias results in biased estimates of calibration parameters \cite{brynjarsdottir2014learning}.

Kennedy and O' Hagan \cite{kennedy2001bayesian} proposed a statistical framework for dealing with various sources of uncertainty in computer models. They introduce a systematic bias function of the variable inputs which links the computer model output with the real process as follows

\begin{equation}
y^R(\mathbf{x}) = y^M(\mathbf{x}, \mathbf{u}^*) + b(\mathbf{x}) ,
\end{equation}

\noindent where $R$ denotes the real process, $M$ is the computer model, $\mathbf{u}^*$ is the true but unknown value of the calibration parameters and $b$ if the bias function. The noisy observed field data are connected to the real process through the formulation

\begin{equation}
y^F(\mathbf{x}) =
y^M(\mathbf{x}, \mathbf{u}^*) + b(\mathbf{x}) + \varepsilon, \text{ where } \varepsilon \sim N(0,\sigma^2_\varepsilon I). 
\label{eq:BC}
\end{equation}

\noindent Considering the potential non-linearity of the bias function a GP prior is used to model the bias. Since the calibration parameters have usually physical meaning, priors for $\mathbf{u}$ are chosen according to knowledge of the underlying science. The computer model is often complex and computationally expensive, therefore KOH replaced it with an emulator which is a statistical approximation to the model. Since models are highly non-linear a flexible Gaussian process (GP) prior is used as an emulator trained on $[\mathbf{x}, \mathbf{u}]-\text{space}$ design.  

Gaussian processes \cite{rasmussen2003gaussian,santner2003design} are a popular choice for computer model emulators due to their data interpolation abilities. Formally, GPs are priors over unknown functions where every finite collection of model outputs $\mathbf{y} $ follows a multivariate Gaussian distribution with a mean function $\boldsymbol{\mu}$ and covariance function $K(\mathbf{x}_i,\mathbf{x}_j).$ A typical choice for the emulator mean is zero or constant and if a prior belief is that the computer model output is a smooth function of inputs, the anisotropic squared-exponential covariance  function 
$K(\mathbf{x}_i,\mathbf{x}_j) = \sigma^2\exp(-\frac{|\mathbf{x}_i-\mathbf{x}_j|^2}{l_j})$
is a common choice. A more flexible alternative is to introduce a new parameter to the covariance function which controls the smoothness of the process at each input direction with  $K(\mathbf{x}_i,\mathbf{x}_j) = \sigma^2\exp(-\frac{|\mathbf{x}_i-\mathbf{x}_j|^{d_j}}{l_j}),$ 
where $1 \leq d_j \leq 2,$ for each $j.$ This is a default emulator choice in this paper.

KOH suggested to infer all unknowns of  \cref{eq:BC} jointly while Bayarri et al. \cite{bayarri2007framework} proposed a modularized framework based on the same formulation.
Instead of estimating all unknown parameters at once, they proceed in two stages. At stage 1 the emulator GP hyper-parameters are estimated and are fixed for stage 2 where the posterior of of the parameters of interest is obtained given the estimated values of stage 1. This method is used in this paper and we are interested to quantify the uncertainty of calibration parameters. For convenience we draw samples for the marginal precision of the bias process and the precision of noise parameter.

Computer model outputs are usually functional and specifically the models we use in this paper produce time series outputs. This creates a computational challenge for the Bayesian calibration approach due to the quadratic complexity $\mathcal{O}(n^3)$ of the GP models. This challenge has been addressed by dimension reduction methods. For example Bayarri et al. \cite{bayarri2007computer} utilized wavelets basis functions while Hidgon et al. \cite{higdon2008computer} 
used singular value decomposition (SVD). When the output curve is relatively simple an approach is to use only some influential time points for the input and output observations.

\section{Bayesian calibration for Windkessel models}
\label{sec:BC_WK}

In this section, we describe our inferential approach to the calibration of WK models. We also outline the experimental case study, which highlights the two main issues that we deal in the real data case, 1) noisy observed data and 2) imperfect computer models. Finally, we briefly describe the data acquisition process for the real case study.  

\subsection{Model and Inference}
\label{sec:BC_WK_Inf}
Windkessel simulators are used to simulate the blood pressure waveform given a set of the calibration parameter values, inflow and time. In the remaining of the paper, we use the WK2 simulator  \cref{eq:WK2} as a computer model choice in the formulation \cref{eq:BC} as it offers interpretable  physical parameters (Compliance, $C$ and Resistance, $R$). \Cref{eq:BC} can be written as follows

\begin{equation}
P^F(t) = P^\textrm{WK2}(I,t,R,C) + b(I,t) + \varepsilon, 
\end{equation}

\noindent where $P^F(t)$ is the observed pressure in time, $P^{WK2}(I,t,R,C)$ is the WK2 pressure simulator, $b(I,t)$ is the model bias and the noise, modeled as $\varepsilon \sim N(0,\sigma_\varepsilon ^2 I).$ 

The WK2 pressure simulator is modeled by a power exponential Gaussian process as described in \Cref{sec:BC}  and is trained on a design of plausible calibration parameter values $R\times C = [0.5,3]^2$. For the bias function $b(I,t)$ we use a GP prior with squared-exponential (or Gaussian) kernel and 0 mean, which means that our prior belief supports a perfect model. Furthermore, we assign prior distributions to the calibration parameters. For both resistance and compliance we choose uniform priors over the range of possible values, that is $R,C\sim Unif(0.5,3).$

We follow the modular approach proposed by Bayarri et al. \cite{bayarri2007framework}. At stage 1 we obtain the smoothness parameters $\mathbf{d} = (d_1,\ldots,d_4),$ the characteristic length scales $\mathbf{l} = (l_1, \ldots, l_4)$ and the marginal precision $\lambda$ of the WK2 emulator by maximizing the marginal log-likelihood and fix them for stage 2. To account for the time dependency we carefully select some  influential points of the output curve and we use time as input along with inflow and the calibration parameters R and C. This choice is reasonable since the output of the WK2 model is smooth without abrupt changes (see \cref{fig:WK2and3}). 

At stage 2 we first run the emulator on the observed (field) data for an initial guess of calibration parameters, $R_0$ and $C_0$. In order to create data for the model bias, we take the difference between the model predictions (runs), $\hat P^\textrm{WK2}(I_{obs},t_{obs},R_0,C_0)$ and the observed data, $P_{obs}$. We use these differences to fit the GP model bias, where we obtain estimates for the model hyper-parameters (marginal precision $\hat{\lambda}^b$, correlation length scales $\hat{l}_1, \hat{l}_2$ and noise precision $\hat{\lambda}^F$) by maximizing the log-marginal likelihood, similar to stage 1. Then following \cite{bayarri2007framework}, we fix the bias correlation length scales to the estimated values and we assume for both precision parameters exponential priors with means equal to a modest multiple of the MLE's. Specifically, the prior distributions are $\lambda^b\sim\Gamma(1,5\hat{\lambda}^b)$ and $\lambda^F\sim\Gamma(1,5\hat{\lambda}^F).$ Finally, to complete the Bayesian formulation we use flat priors on the range of possible values for both calibration parameters with $R\sim Unif(0.5,3)$ and $C\sim Unif(0.5,3).$

The standard approaches for obtaining physical parameters estimates for the WK models. A benefit of Bayesian calibration is that quantifies the parameters uncertainty. For example conventional approaches are solving the optimization problem $\text{argmin}_{R,C}RSS$ (where $RSS = \sum_{i=1}^{n}(P^\textrm{WK}(t_i) - P^{obs}(t_i))^2, i=1,\dots n,$ $P^\textrm{WK}$ is the predicted values of the $WK$ model at the corresponding observed inflow points and $P^{obs}(t_i)$ denotes the observed pressure at temporal locations $t_i$) using non-linear least squares and produce point estimates, while Bayesian calibration offers the bias-corrected posterior distribution $P(R,C\mid P(t), I(t)).$

\begin{figure}[h]
	\includegraphics[width=1\columnwidth]{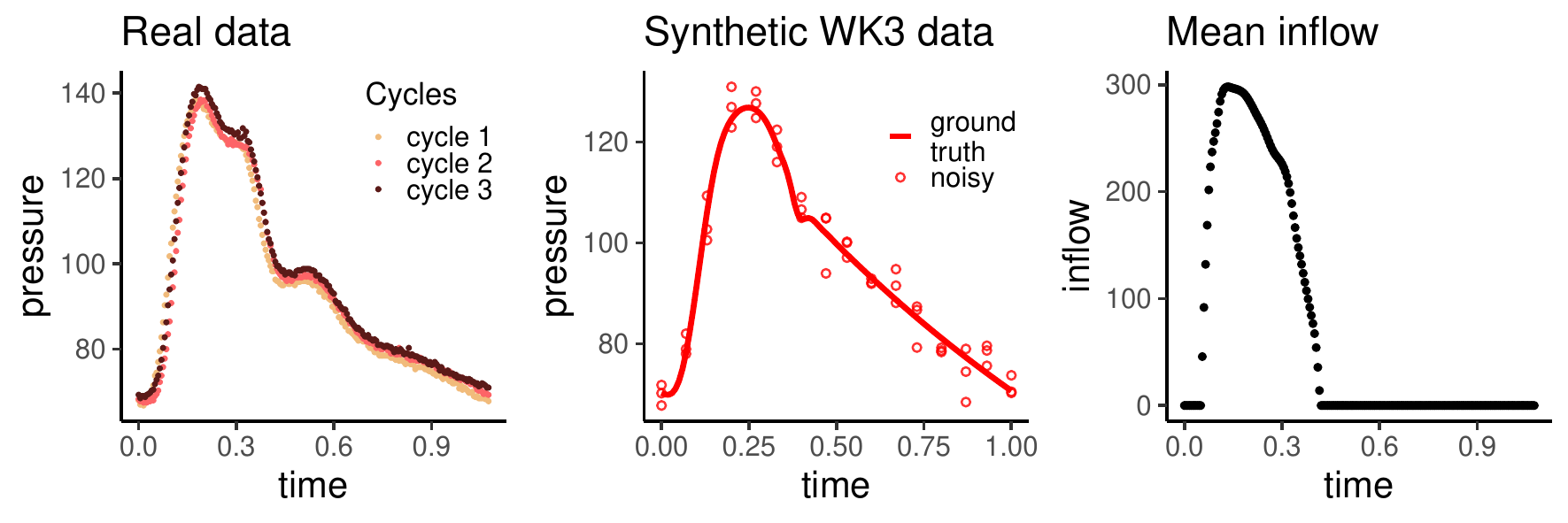}
	\caption{The 3 cycles of blood pressure for the real case study (left). Noisy WK3 simulated data of the synthetic example (middle) and the observed mean inflow (right)}
	\label{fig:RealSynthInflow}
	
\end{figure}

\subsection{Synthetic case study, WK3 model as ground truth}
\label{sec:BC_WK_Synth}
The aim of the synthetic case study is two folded: i) To explore the properties of the standard approach for inference (minimizing the the RSS as described in \Cref{sec:BC_WK_Inf}), when the model considered is biased, and 2) To study the performance of Bayesian calibration. To validate our modeling approach we need synthetic data from a model where the true values of the physical parameters are known. A broadly used model in the literature is the WK3 model due to its ability to fit the observed data reasonably well. We simulate data from this more complex model and the main goal is to show that our modeling approach (3) (WK2 + model bias) can learn the model parameters and reconstruct the blood pressure waveform. In terms of interpretability, the difference between our modeling approach and the WK3 model is that the vascular resistance of the WK2 model is $R$ while in WK3 equals $R_1+R_2.$ Furthermore, the total arterial compliance values, $C$ should be identical in both cases.

We consider four different setups of WK3 model-simulating data for four different sets of model parameters $R_1,R_2 \text{ and } C$  (see \cref{table:synthetic}). In two of the setups the total arterial compliances are equal ($C = 0.8$, see \cref{table:synthetic}, Setup1 and Setup2) and two other cases where the vascular resistances are equal ($R=1.1$ see \cref{table:synthetic}, Setup1 and Setup3). We have also considered a case where the parameter values are more distant in the parameter space ($C=1,R=1.42$, see \cref{table:synthetic}). The aim of this experimental Setup is to see if we can identify smaller (Setups 1-3) or larger (Setup 4) changes in the parameters, for example after a period of medication and/or lifestyle intervention.  Since pressure is measured by sensors,  data are usually noisy and for this reason, we add artificial independent and identically distributed (i.i.d.) noise as $\varepsilon_i\sim N(0, 4^2).$ In a real case scenario, it is easy to obtain many blood pressure cycles as can be seen in \cref{fig:RealSynthInflow} with a constant temporal resolution  $\approx 0.05$ seconds. 

We begin with a simulation study which validates the accuracy of optimization using the WK2 and WK3 models. It is common that the blood pressure measurements are noisy and this can produce bias in the parameter estimates. For each of the setups of  \cref{table:synthetic} we produce 100 noisy synthetic data sets, by simulating the blood pressure waveform and adding i.i.d. $N(0, 4^2)$ noise. For each of these simulated scenarios, parameters are estimated using non-linear least squares and we report the average parameter values of the 100 under the 100 noisy simulations along with $90\%$ empirical confidence intervals.

In the box-plots of \cref{fig:sythnSim} we summarize the results of the simulation study regarding the convergence of the optimization approaches for calibration of the windkessel models. For each of the setups, we have simulated 100 noisy blood pressure waveforms from the WK3 model by using the same inflow and adding i.i.d. $N(0,4^2)$ noise to the pressure output. Then we have optimized the RSS of the WK2 and the WK3 models by using non-linear least squares as described in \Cref{sec:BC_WK_Inf}).

\subsection{Real case study}
\label{sec:BC_WK_Real}
Real case study data was obtained from a pilot randomized controlled trial conducted by researchers at the Department of Circulation and Medical Imaging at the Norwegian University of Science (NTNU) and Technology and St. Olavs University Hospital in Trondheim, Norway. Secondary aims of the trial include contribution to the development and improvement of biomechanical and computer models that describe the underlying hemodynamics of high blood pressure. The trial was approved by the regional medical ethics-committee (REK 2019/1084) and registered on clinicaltrials.org (Identifier: NCT 04151537).

Two subjects, 45-64 years of age, with high blood pressure (without prescribed blood pressure medication, diagnosed diabetes or cardiovascular disease) and complete records on brachial pressure and aortic blood inflow at baseline were selected for the real case study.

Continuous (beat-to-beat) brachial pressure (see  \cref{fig:RealSynthInflow}, left) was assessed with Finometer PRO (Finapres Medical Systems, Enschede, Netherlands) on the right arm while the subject rested on their left side in the supine position, thereby enabling simultaneous and synchronized assessment of aortic blood inflow. Blood inflow (see  \cref{fig:RealSynthInflow}, right) was assessed using Doppler flow and diameter measurements in the left ventricular outflow tract of the aorta. All echocardiographic assessments were performed by an experienced sonographer using a Vivid e95 scanner (GE Vingmed Ultrasound, Horten, Norway) with a 4VC phased array three-dimensional transducer and followed international recommendations \cite{mitchell2019guidelines}.

The trial is part of the My Medical Digital Twin (MyMDT) project, which is funded by NTNU through the Digital Transformation Initiative.

\section{Results}
\label{sec:res}
In this Section we present the results for the synthetic and real case studies. The synthetic case study is divided in two simulation studies. The first study evaluates the accuracy of the optimization approaches by exploring the convergence of the mean value and the coverage of the empirical confidence intervals. In the second synthetic study we use the WK3 model as ground truth from which we simulate noisy blood pressure data. Then we apply the Bayesian Calibration approach as detailed in  \Cref{sec:BC_WK_Inf} and \ref{sec:BC_WK_Synth} and we compare it with the optimization to the WK2 and WK3 models. Finally, we apply the same methods to the real data obtained by two individuals as described in \Cref{sec:BC_WK_Real}.

\begin{figure}[h!]
	\centering
	\includegraphics[width=8cm, height=6cm]{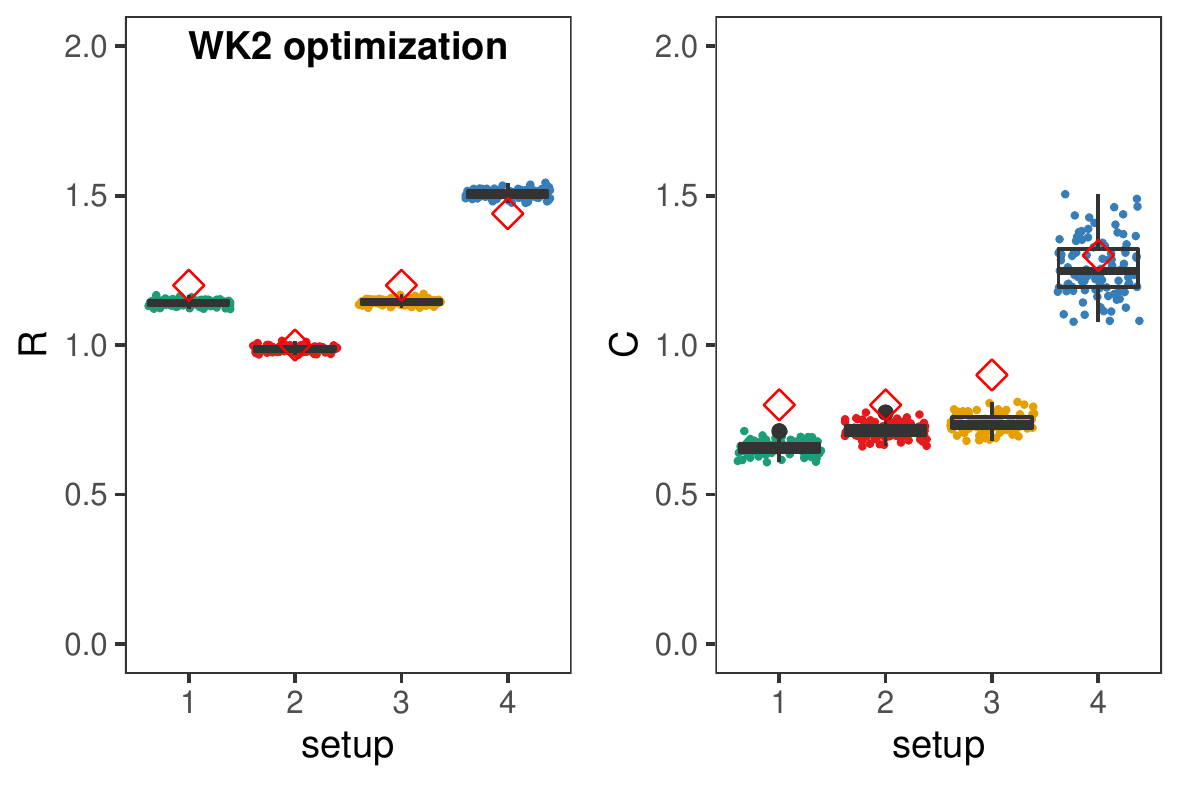} %
	\centering
	\includegraphics[width=13cm, height=6cm]{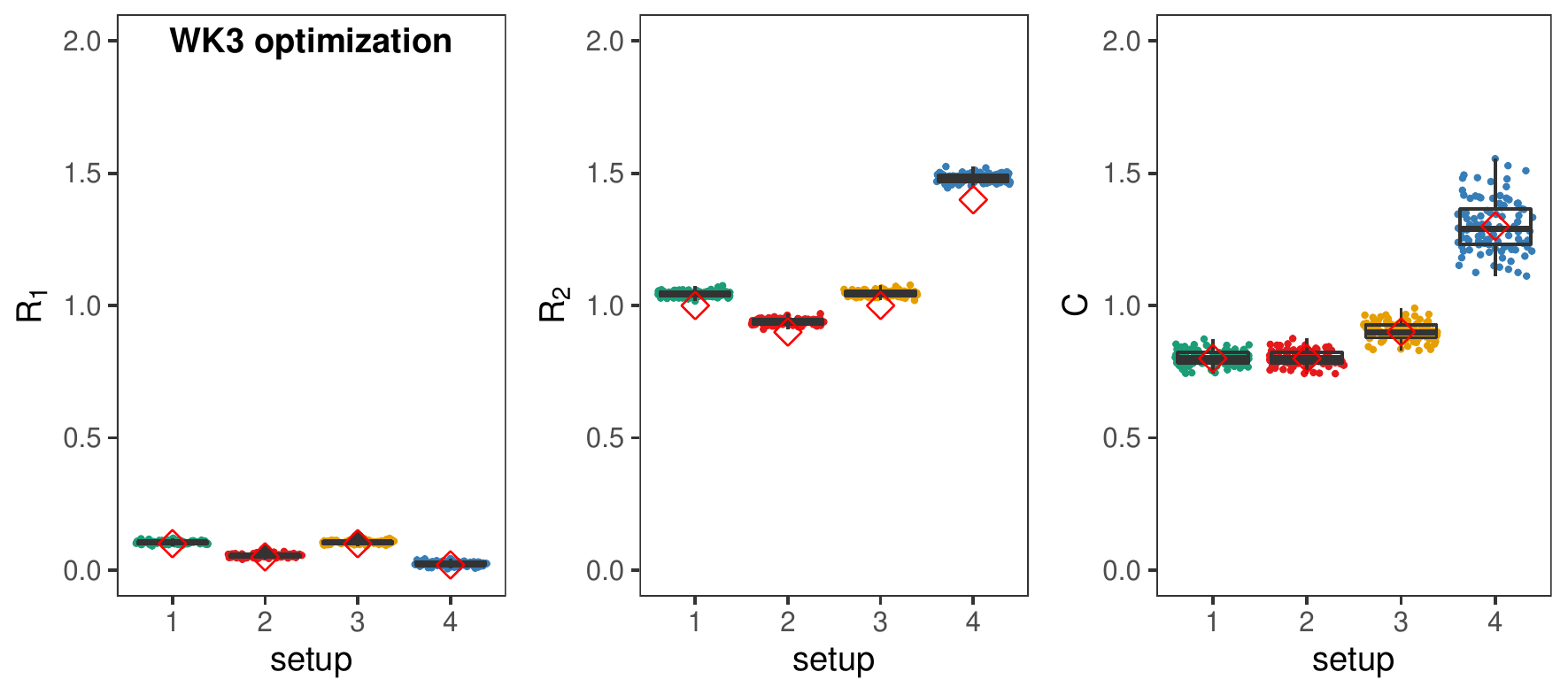} %
	\caption{Simulated data from WK3 mode with added i.i.d. $N(0,4^2)$ noise for the same parameter setups as in Table \ref*{table:synthetic}. 
		The first row corresponds to the WK2 model and the second corresponds to the WK3. The red diamond is the true value for each parameter. 
		For the WK2 model $R$ represents the vascular resistance and its value equals $R_1 + R_2.$}
	\label{fig:sythnSim}%
\end{figure}

\begin{table}[h!]
	\begin{center}
		\scalebox{0.8}{
			\begin{tabular}{|c |c |c |c |c |c |c |}
				\hline
				&Parameters & Real 
				&\multicolumn{2}{|c|}{WK2 optim}
				&\multicolumn{2}{|c|}{WK3 optim}\\
				\cline{4-7}
				& & 
				&mean & $90\%$ CI
				&mean & $90\%$ CI\\
				\cline{2-7}
				Setup 1
				&$C$  &0.8
				&0.657 &$(0.620,0.693)$
				&0.802 &$(0.757,0.850)$\\
				
				&$R (=R_1 + R_2)$  &1.1 
				&1.140 & $(1.124,1.153)$
				&1.149 &$(1.125,1.173)$\\
				
				&$R_1$  &0.1
				&- &-
				&0.105 &$(0.097,0.113)$\\
				
				&$R_2$  &1
				&- &-
				&1.044 &$(1.028,1.060)$\\
				\hline
				
				Setup 2
				&$C$  &0.8 
				&0.715 & $(0.674,0.754)$
				&0.801 &$(0.756,0.849)$\\
				
				&$R (=R_1 + R_2)$  &0.95 
				&0.987 &$(0.971,1.000)$
				&0.993 &$(0.969, 1.016)$\\
				
				&$R_1$  &0.05 
				&- &-
				&0.055 &$(0.046,0.063)$\\
				
				&$R_2$  &0.9 
				&- &-
				&0.938 &$(0.923,0.953)$\\
				\hline
				
				Setup 3
				&$C$  &0.9  
				&0.740 &$(0.694,0.785)$
				&0.901 &$(0.846,0.961)$\\
				
				&$R (=R_1 + R_2)$  &1.1 
				&1.144 &$(1.128,1.157)$
				&1.152 &$(1.127,1.175)$\\
				
				&$R_1$  &0.1 
				&- &-
				&0.105 &$(0.096,0.113)$\\
				
				&$R_2$  &1 
				&- &-
				&1.047 &$(1.031,1.062)$\\
				\hline
				
				Setup 4
				&$C$  &1.3 
				&1.261 &$(1.112,1.434)$
				&1.300 &$(1.15,1.482)$\\
				
				&$R (=R_1 + R_2)$  &1.42 
				&1.505 & $(1.483,1.524)$
				&1.505 &$(1.469 1.539)$\\
				
				&$R_1$  &0.02 
				&- &-
				&0.024 &$(0.011,0.036)$\\
				
				&$R_2$  &1.4  
				&- &-
				&1.481 &$(1.458,1.503)$\\
				\hline

			\end{tabular}
		}
		\caption{Results for the simulation on  optimization methods using the WK3 model as ground truth. For each parameter setup we present the mean estimate of the 100 simulations and the $90\%$ empirical confidence intervals.}
		\label{table:simulation100}
	\end{center}
	
\end{table}

\subsection{Results, synthetic case with the WK3 model as ground truth}
\label{sec:Res_synth}
\subsubsection{Simulation study for the properties of optimization approaches}

From \cref{fig:sythnSim} we observe that the WK2 model optimization systematically underestimates $C$ and for setups $1-3$  the $90\%$ empirical confidence intervals do not cover the true value (see also \cref{table:simulation100}). For the same setups the mean of the estimated $C$ values obtained by optimizing the WK3 model, converge to its true values. For setup 4 we have simulated data from a low amplitude curve (see  \cref{fig:PredSynth}, top plot, blue waveform) which corresponds to a non-hypertensive individual (systolic blood pressure $\approx 120$ and diastolic $\approx$ 80). For this setup we observe that the variability of the estimated $C$ is quite large. This is the case for both WK2 and WK3 and suggests that the uncertainty of these estimates should be taken into account. 

As described in \Cref{sec:WK} the vascular resistance is given as $R$ for the WK2 mode, while for the WK3 model equals to $R_1 + R_2.$ From  \cref{fig:sythnSim} we observe that for all setups $R$ is slightly biased and the true value of the parameter does not fall in the $90\%$ empirical confidence interval. These results also holds for the WK3 model. We see that the estimates of the parameter $R_2$ are systematically biased downwards and again the true $R_2$ values do not fall in the $90\%$ empirical confidence intervals. This might reveal an identifiability problem for the WK3 model.

To summarize, the WK2 model produce biased estimates of the parameters ($R$ and $C$) if the truth is the WK3 model. The WK3 model where we used as ground truth, suggest that estimated $C$ value can vary considerable. Hence, we should account for the uncertainty of the estimates and the estimated $R_2$ parameter seems to be non-identifiable. 


\begin{figure}[htbp]
	\centering
	\includegraphics[width=1\textwidth]{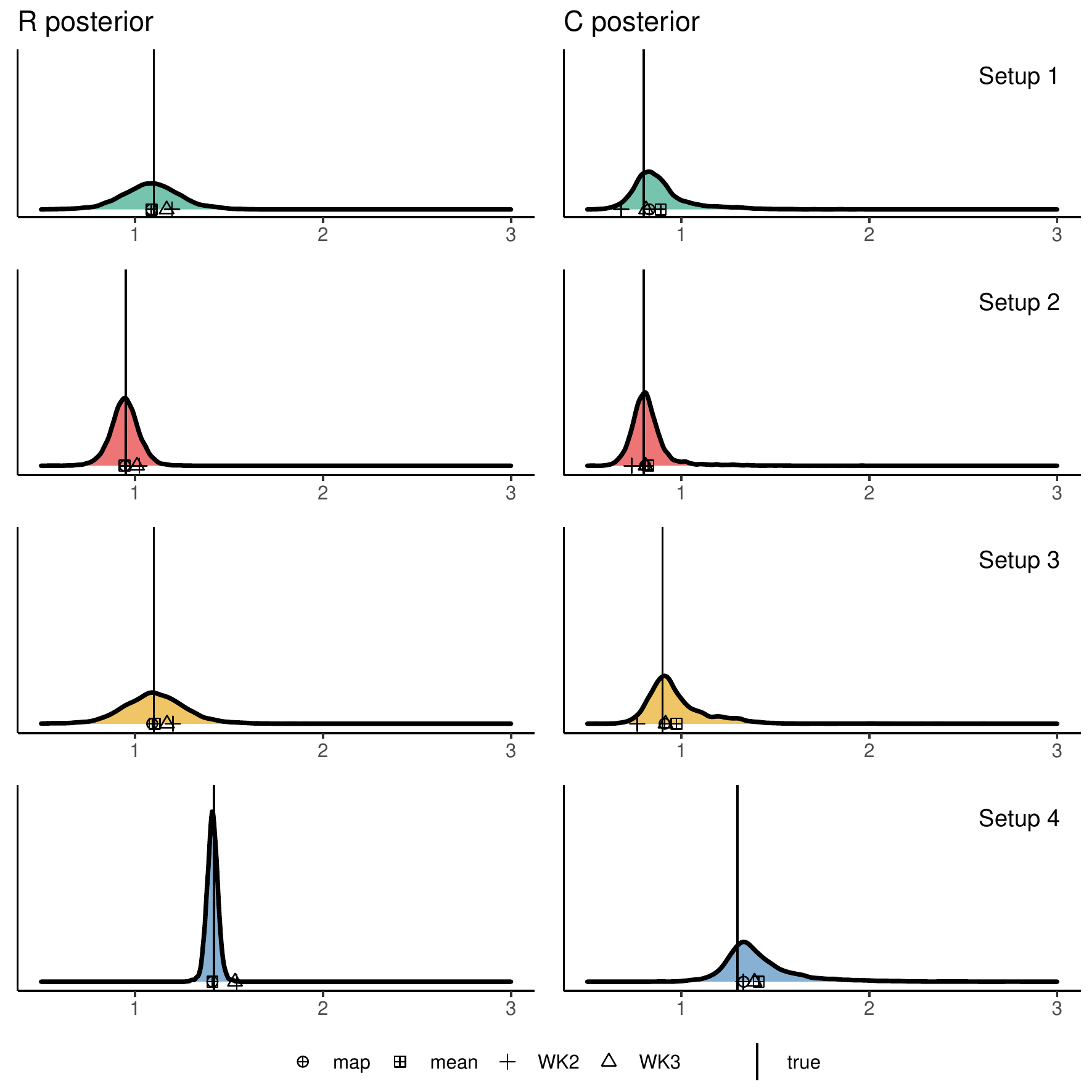}
	\caption{Marginal posterior distributions of the calibration parameters R and C for all 4 experimental setups. Mean and map estimates are provided for the Bayesian calibration along with point estimates obtained by minimizing the distance between the observed pressure curve and the WK2-WK3 models using optimization.}
	\label{fig:RCmarg}%
\end{figure}

\subsubsection{Bayesian calibration with the WK3 model as ground truth}

\begin{figure}[htbp]
	\centering
	\includegraphics[width=10cm, height=5cm]{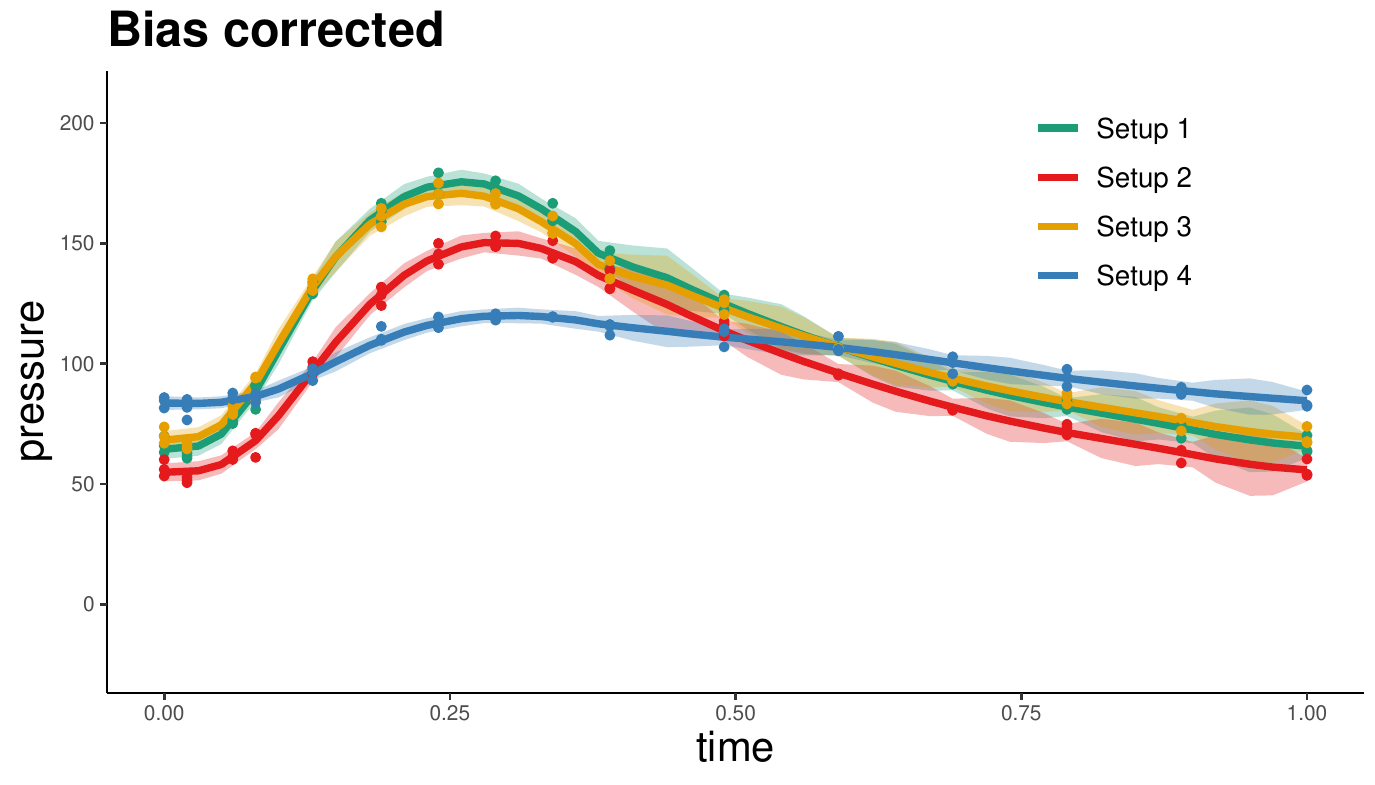} %
	\centering
	\includegraphics[width=12cm, height=4cm]{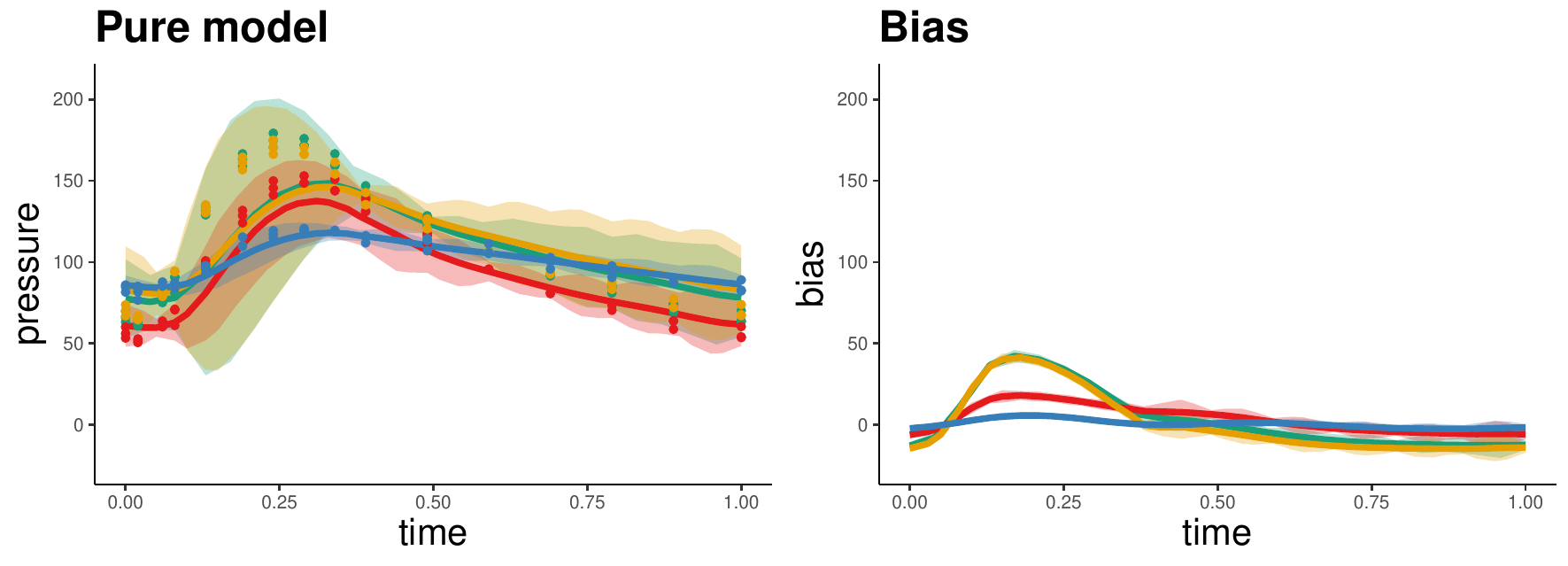} %
	\caption{Bias-corrected predictions for out of sample data points (top). Pure model preditions (bottom-left) and functional bias (bottom-right). The solid lines are the posterior means and the shaded areas represent the $90\%$ prediction intervals. The dots are the observed data for 3 cycles.}
	\label{fig:PredSynth}%
\end{figure}

For this synthetic study we simulated data from the WK3 model and added noise as before, $\varepsilon \sim N(0,\sigma^2_\varepsilon I)$ , but now we have considered a realistic scenario where we have obtained only three cycles for each setup (for Example see \cref{fig:RealSynthInflow}, second plot or see  \cref{fig:PredSynth}). In \cref{fig:RCmarg} we plot the marginal posterior distributions of the vascular resistance, $R$ and total arterial compliance, $C$ along with the maximum a posteriori probability estimate (map), that is the point where the probability mass is concentrated and by averaging all over the uncertainties we produce the mean estimates as well for the Bayesian Calibration approach. To compare with conventional approaches we also estimated parameters using optimization for the WK2 and WK3 models.

\begin{figure}[hthb]
	\centering
	\includegraphics[width=9cm, height=8cm]{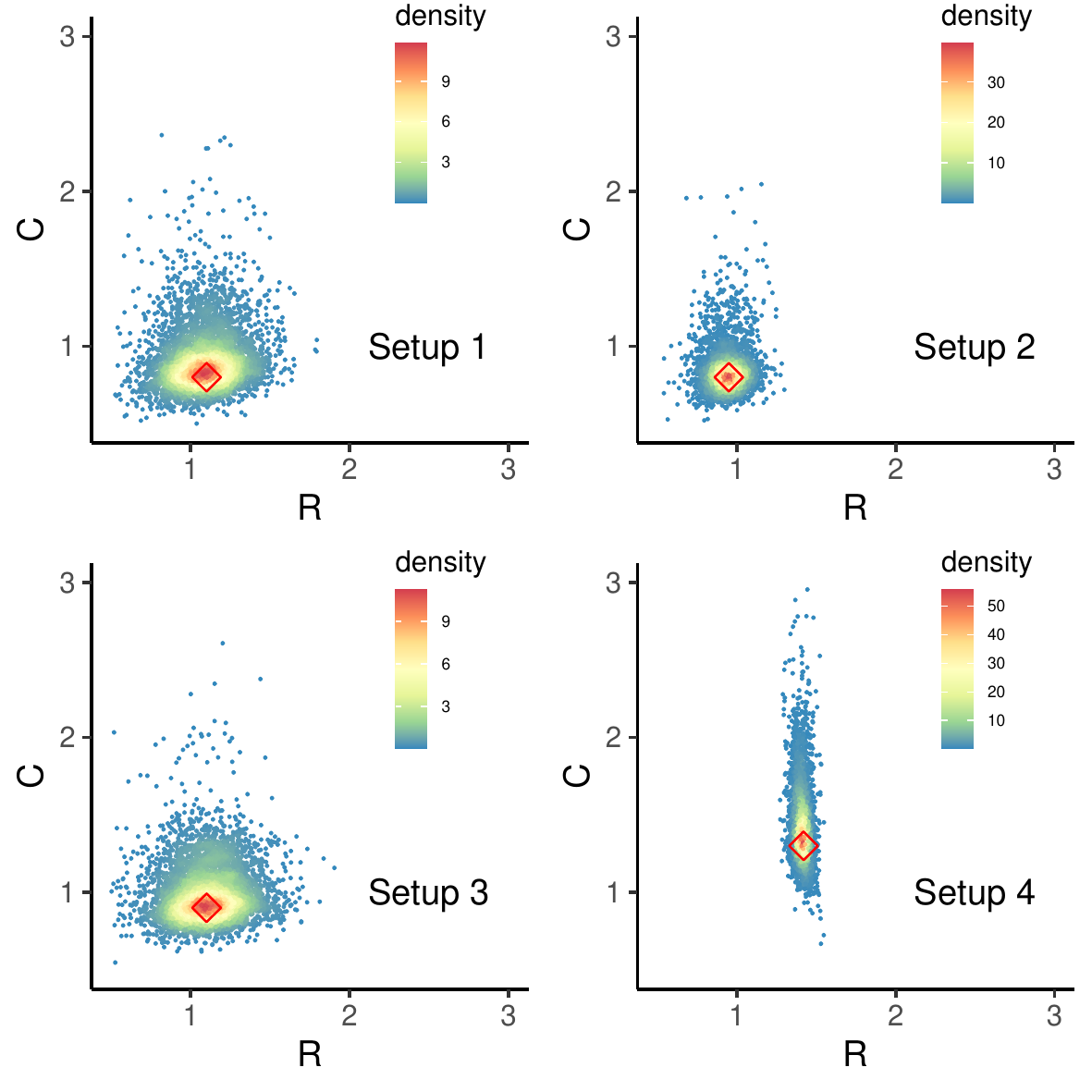}
	\caption{Joint posterior distributions of the calibration parameters R and C for the four synthetic setups. The red diamond is true value of the WK3 model, where the data were simulated.}
	\label{fig:Joint}%
\end{figure}

As mentioned in \Cref{sec:WK} the value of vascular resistance $R$ in WK2 model corresponds to $R_1 + R_2$ for the WK3 model and this property holds for Bayesian calibration as can be seen in  \cref{fig:RCmarg} and in \cref{table:synthetic}. For example, at setup 1 we have $R_1 + R_2 = 0.1 +1 = 1.1$ and as can be seen in \cref{fig:RCmarg}, the posterior is approximately symmetric around the mean, which is 1.087 (\cref{table:synthetic}). 
We see in \cref{table:synthetic} by optimizing the WK2 model, underestimated compliance estimates for the setups 1-3 where $R_1\geq 0.05$ (see also the simulation study, \cref{fig:sythnSim} and the \cref{table:simulation100}). This bias increases with the value of parameter $R_1,$ while for setup 4 where $R_1 = 0.02$ and WK3$\to$WK2 the estimated compliance is much closer to the truth value. By optimizing the WK3 model (ground truth) we obtain good estimates of C, but R seems to be systematically overestimated (see \cref{table:synthetic}) something we have found by simulations as well (see \cref{fig:sythnSim}). For the Bayesian Calibration setting, we see that estimated (posterior mean and map) compliance, $C$ are closer to the true values for all four setups compared to optimizing the WK2 model (see \cref{table:synthetic} an \cref{fig:RCmarg}). Furthermore, mean and map estimates are quite similar for all 4 different parameter configurations and as can be seen in \cref{fig:RCmarg} the marginal posteriors for both R and C are concentrated around the true values. In \cref{fig:Joint} we also see the joint posterior distributions for $R$ and $C$ plotted by using 7000 thinned MCMC samples. We see that in all four setups the joint posterior is concentrated at the true $(R,C)$ value (red diamond).

In \cref{fig:PredSynth} are the Bias-corrected model predictions, the pure-model predictions, that is the WK2 model fitted at the estimated $R$ and $C$ parameter values (from Bayesian calibration) but without discrepancy (or simply the WK2 emulator without the discrepancy) and the (functional) bias for each setup. We observe that with Bayesian calibration we can recover the observed data, we find that the model itself (pure-model plot) can not achieve. The WK2 model can not capture blood pressure waveforms with high amplitude especially during diastole ($t \approx 0.3)$ and this comes with quite large uncertainty in the model predictions (see pure model plot,  shaded areas represent the 90\% prediction intervals). Larger values of the $R_1$ parameter result in larger bias (\cref{fig:PredSynth}, Bias plot) and larger uncertainty for pure model predictions. For the setup 1 (where $R_1=0.1$),  pure model predictions are highly uncertain, while for setup 4 (where $R_1 = 0.02$) pure model prediction is quite close to the data. This is also depicted to the estimation of $R$ where the posterior for setup 4 is much narrower (see \cref{fig:RCmarg}) with the $90\%$ credible interval $(1.363, 1.460),$ while for setup 1 the interval is $(0.837, 1.348)$ (see \cref{table:synthetic}).

A desired ability of the Windkessel models is to capture changes in the physical parameters after some period of medication or physical exercise. For example we would expect after a period of medication to observe a decrease in the total arterial compliance, $C$ which is inversely related to the elasticity. The posterior distributions offer a degree of uncertainty for these changes. Here, we kept some of the parameters of different configuration identical, in order to investigate if we are able to identify possible parameter changes, even if they are very small. For example, setup 1 and setup 3 have both true resistance value $R=1.1$ (see \cref{table:synthetic}). We see that their marginal posteriors for $R$ seem to be almost identical, while the posterior of the setup 4 with real value $1.3$ most of the probability mass is far from $1.1.$ For $C$, setup 1 and 2 have a true value of $0.8,$ though we observe that for the first the credible interval is almost two times larger. This stems from the fact that bias for setup 1 is larger (see \cref{fig:PredSynth}). For small changes of $C$, we observe that for setups 1 and 3 ($C_{real_1} = 0.8$ and $C_{real_3} = 0.9$) although map estimates are very close to the true values the corresponding posteriors are relatively wide and $90\%$ credible intervals are almost identical (see \cref{table:synthetic}).

\begin{table}[htbp]
	\begin{center}
		\scalebox{0.8}{
			\begin{tabular}{|c|c| c |c |c |c |c |c|}
				\hline
				&Parameters & Real 
				&WK2 optim
				&WK3 optim
				&\multicolumn{3}{|c|}{Bayes Calib} \\
				\cline{6-8}
				& & & &
				&map &mean & $90\%$ CIs \\
				\cline{2-8}
				Setup 1
				&C 
				& 0.8 &0.656 &0.793 
				&0.830 & 0.890 & (0.705, 1.227)\\
				\cline{2-8}
				&$R(=R_1+R_2)$
				&1.1
				&1.192 &1.163 &1.087 
				&1.087  & (0.837, 1.348)\\
				
				&$R_1$
				& 0.1 
				&- &0.106
				&- &-  &-\\
				
				&$R_2$
				&1
				&- &1.056
				&- &-  &-\\
				\hline
				Setup 2
				&C 
				&0.8 &0.709 &0.791  
				&0.808 &0.824 & (0.702, 0.995)\\
				\cline{2-8}
				&$R(=R_1+R_2)$ 
				& 0.95  
				&1.019  &1.005 
				&0.946  &0.945  & (0.820, 1.067)\\
				
				&$R_1$ 
				&0.05 
				&-  &0.055
				&- &-  &-\\
				
				&$R_2$ 
				&0.9  
				&-  &0.95
				&- &-  &-\\
				\hline
				Setup 3
				&C & 0.9  
				&0.736 &0.892 
				&0.911 &0.974 & (0.791, 1.293)\\
				\cline{2-8}
				&$R(=R_1+R_2)$  
				& 1.1 
				&1.197  &1.165 
				&1.091 & 1.104 & (0.841, 1.379)\\
				&$R_1$
				& 0.1
				&-  &0.106 
				&- &-  &-\\
				&$R_2$ 
				&1 
				&-  &1.059
				&- &-  &-\\
				\hline
				Setup 4
				&C 
				&1.3 &1.338 &1.389 
				&1.332 &1.412 & (1.183, 1.804)\\
				\cline{2-8}
				&$R(=R_1+R_2)$ 
				& 1.42
				&1.531  
				&1.523
				&1.412 & 1.412 & (1.363, 1.460)\\
				&$R_1$ 
				& 0.02
				&-  &0.023
				&- &-  &-\\
				&$R_2$ 
				& 1.4
				&-  &1.5
				&- &-  &-\\
				\hline
				
			\end{tabular}
		}
		\caption{Synthetic data results. As mentioned in \Cref{sec:WK} the vascular resistance $R$ for the WK3 model is given as $R_1 + R_2.$
			Since we have 3 noisy blood pressure cycles parameter estimates differ slightly and for this reason we present the average estimate for each parameter. In the Appendix there is a more detailed table with parameter estimates for each cycle as well.}
		\label{table:synthetic}
	\end{center}
	
\end{table}

\begin{figure}[htbp]
	\centering
	\includegraphics[width=1\textwidth]{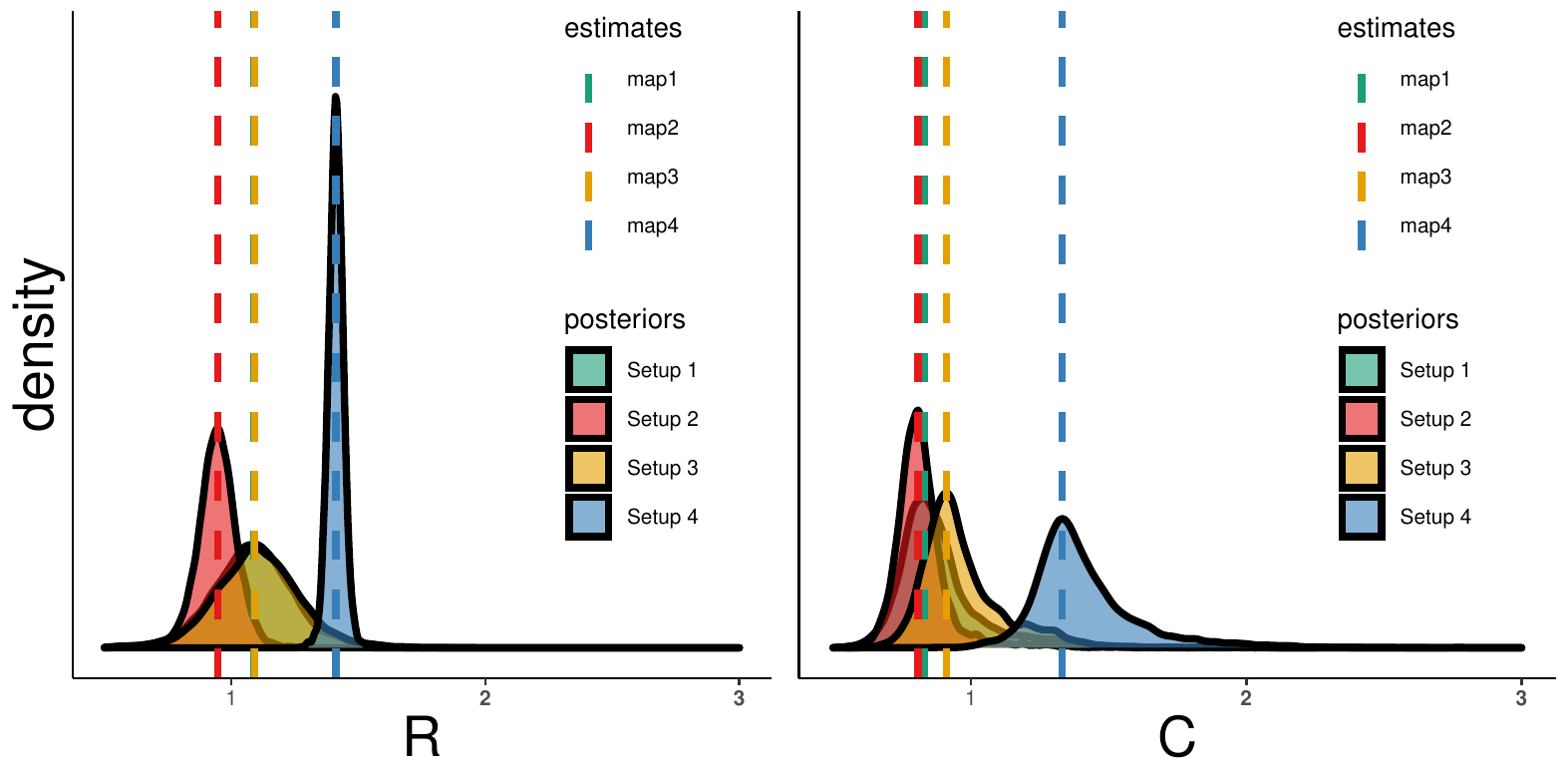}
	\vspace{-0.3cm}
	\caption{Marginal Posterior densities for the physical parameters total arterial compliance-C and vascular resistance-R. The dashed vertical lines are the map estimates for each setup (see also Table \ref{table:synthetic}).}
	\label{fig:postRC}
\end{figure}

\subsection{Real case study results}

\begin{figure}[hthb]
	\centering
	\includegraphics[width=6cm, height=5.5cm]{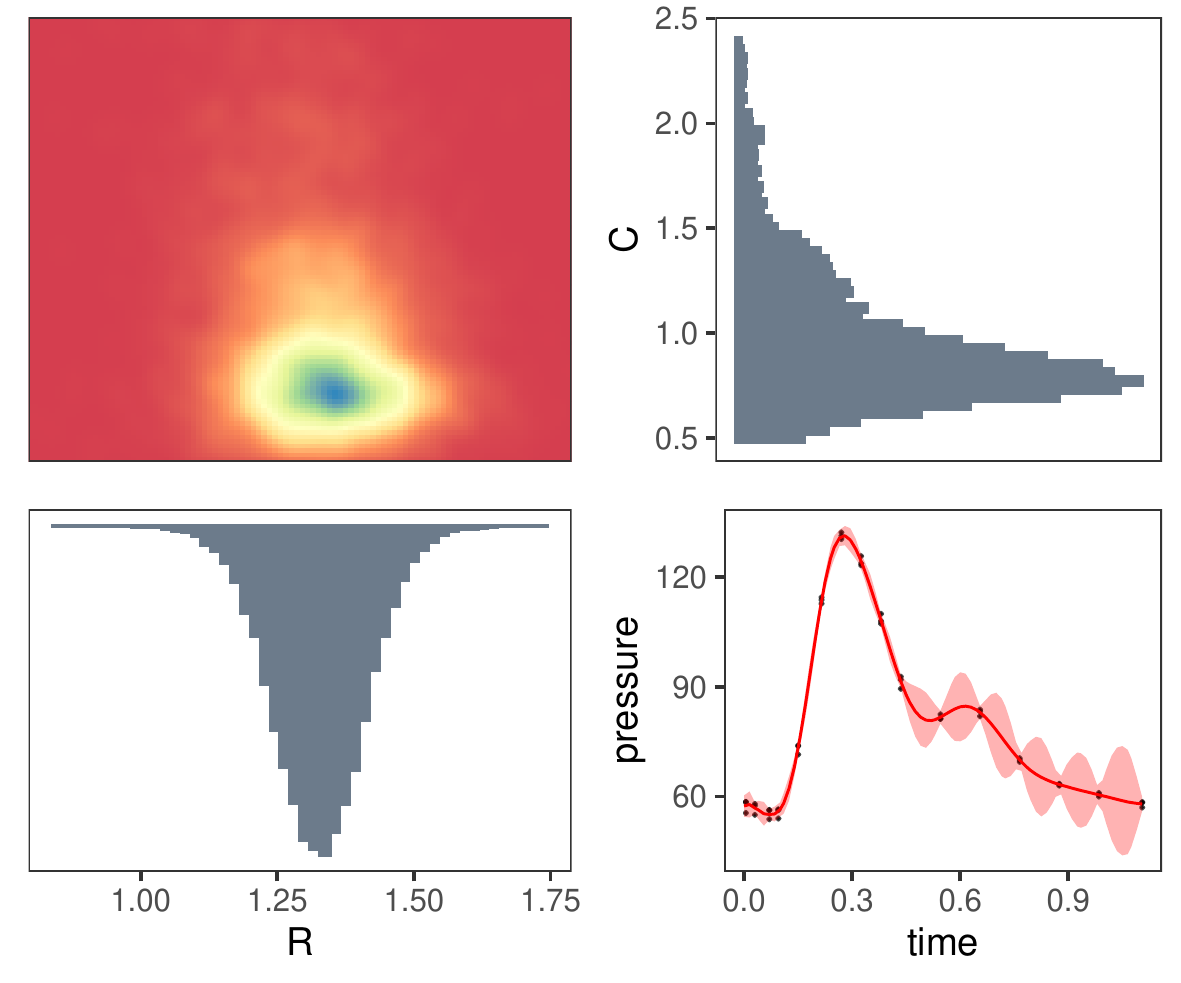} %
	\qquad
	\includegraphics[width=6cm, height=5.5cm]{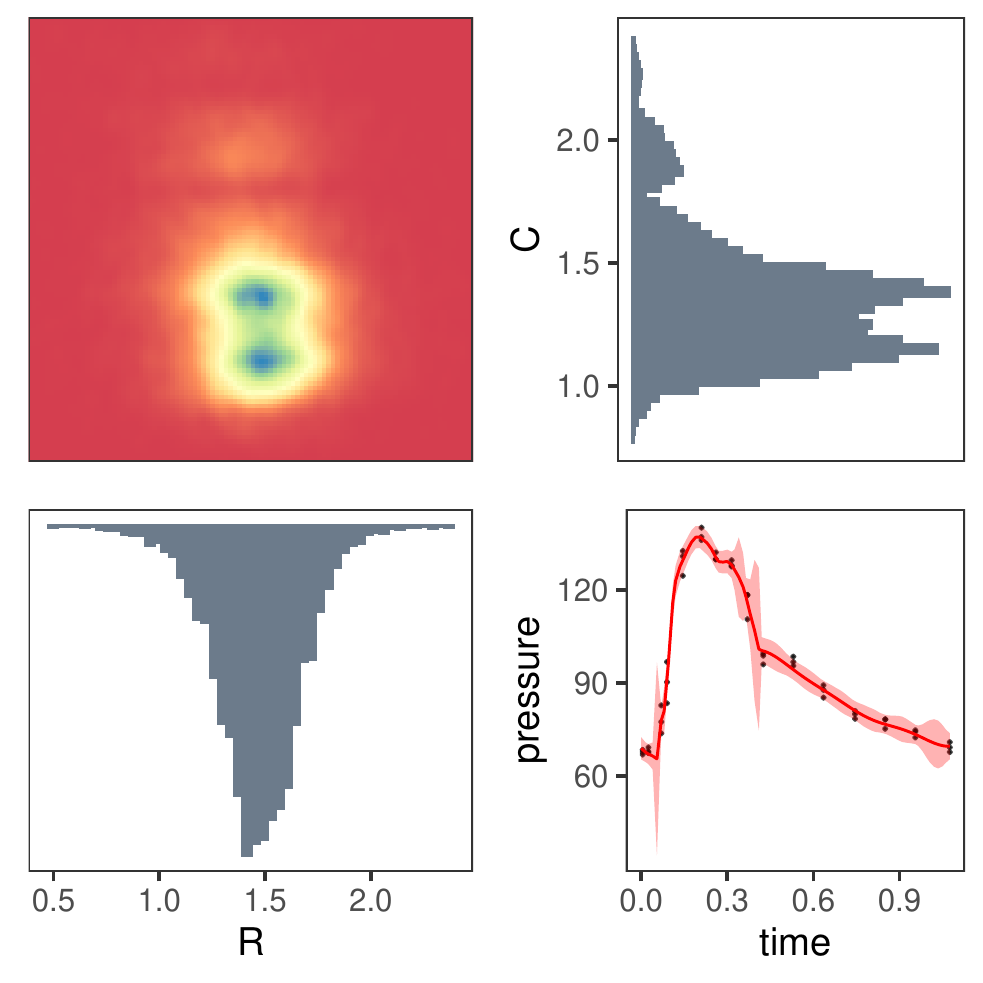} %
	\caption{Real case study with personal data from two individuals. For each person we draw images of the joint posterior density of R and C (top-left). 
		The corresponding marginal posteriors are represented by the histograms in the off-diagonal. In the bottom-right we plot the corresponding model mean predictions (solid red lines)
		with the $90\%$ prediction intervals (shaded areas) from Bayesian calibration and the observed data points (black-dots).}%
	\label{fig:RealCase}%
\end{figure}

We have processed three blood pressure cycles for two individuals (see \cref{fig:RealCase}) and we applied the three methods (Bayesian calibration, optimization to WK2 and to WK3) as in \Cref{sec:Res_synth}. For each of the optimization methods (WK2 and WK3) we obtained estimates of the physical parameters R and C for all three cycles (see \cref{table:RealCase}).  For the Bayesian calibration we have synchronized the three cycles in the same time interval (see \cref{fig:RealSynthInflow}, left plot). 

For the first individual (Case 1 in \cref{table:RealCase}) noise in the data seems negligible (see \cref{fig:RealSynthInflow}) and consequently C and R estimates using WK2 and WK3 model are very similar for all cycles. Both WK2 and WK3 models yield a compliance estimate $\approx 0.7.$ This stems from the fact that $R_1\to 0$ and consequently 
WK3$\to$WK2. Bayesian calibration map estimate of compliance is slightly larger (0.775) while averaging all over the uncertainties the mean estimate is larger ($=0.979$) than the map estimate and the corresponding $90\%$ credible interval is (0.592, 1.727). Regarding the vascular resistance, R all three approaches yield similar estimates $\approx 1.3$ and the $(90\%)$ credible interval of Bayesian calibration is significantly narrower (0.838, 1.339) than the total arterial compliance.

The results for the second individual are quite different. First we observe in \cref{fig:RealCase} that both individuals have relatively similar values of systolic and diastolic blood pressures. Compared to the first individual, now the noise in the data seems significantly larger (see \cref{fig:RealSynthInflow}, left). This creates a cycle to cycle variability in the parameters estimate. In \cref{table:RealCase} we see that for WK2 optimization, the first cycle estimate is larger than the other two cycles and the same holds for WK3 optimization. However, in this case WK2 and WK3 optimization estimated compliance values differ, where WK3 produces considerably larger estimates. Interestingly, Bayesian calibration (see \cref{fig:RealCase}, first row-right) reveals similar behavior with a bimodal marginal posterior of C. As in the other two approaches the modes have similar distance ($\approx 2$), while the modes or the mean estimates are a compromise between WK2 and WK3 optimization estimates. Furthermore, the estimated R for both optimization methods are about $1.3$ for all cycles, while Bayesian calibration mean and map estimates coincide due to the approximately symmetric marginal for R.

\begin{table}[hthb]
	\begin{center}
		\scalebox{0.6}{
			\begin{tabular}{|c |c|c |c |c |c |c | c |c |c |c|c|c|}
				\hline
				
				& Parameters
				&\multicolumn{4}{|c|}{WK2 optim} 
				&\multicolumn{4}{|c|}{WK3 optim} 
				&\multicolumn{3}{|c|}{Bayes Calib} \\
				\cline{3-13}
				&
				&cycle 1 &cycle 2 &cycle 3 &mean 
				&cycle 1 &cycle 2 &cycle 3 &mean  
				&map/modes &mean & $90\%$ CIs\\
				\cline{2-13}
				
				
				&C  
				&0.695 &0.670 &0.685 &0.683 %
				&0.704 &0.682 &0.703 &0.696 %
				&0.775 &0.979 & (0.592, 1.727)\\%
				\cline{2-13}
				Case 1
				&$R (=R_1+R_2)$
				&1.306 &1.323 &1.308 &1.312 %
				&1.306 &1.324 &1.309 &1.313
				&1.328 &1.326  & (0.838, 1.339)\\ %
				&
				$R_1$
				&- &- &- &- %
				&0.008 &0.012 &0.016 &0.012 
				&- &- &-\\
				&
				$R_2$
				&- &- &- &- 
				&1.298 &1.312 &1.294 &1.301
				&- &- &-\\
				\hline
				&C  
				&1.215 &1.041 &1.060 &1.105 %
				&1.644 &1.347 &1.400 &1.464 %
				&(1.149,1.386) &1.352 & (1.032, 1.907)\\
				\cline{2-13}
				Case 2
				&$R( = R_1 +R_2)$ 
				&1.279 &1.284 &1.307 &1.290 %
				&1.283 &1.288 &1.311 &1.294 
				&1.463 &1.466 &(1.135, 1.779)\\ %
				\cline{3-13}
				&$R_1$
				&- &- &- &- %
				&0.167 &0.145 &0.016 &0.157
				&- &- &-\\ %
				\cline{3-13}
				&$R_2$
				&- &- &- &- %
				&1.116 &1.143 &1.152 &1.137  
				&- &- &-\\ %
				\hline
				
			\end{tabular}
		}
		\caption{Physical parameters estimation results. Three methods are presented: WK2/WK3 model optimization and Bayesian calibration. For the optimization methods , the parameter estimates are reported for all three cycles along with the corresponding means. The  WK3 parameters $R_1$ and $R_2$ are reported as well. At Case 2, the marginal distribution of C is bimodal and the corresponding mode values are reported.}
		\label{table:RealCase}%
	\end{center}
\end{table}

\section{Conclusions}
\label{sec:conclusions}

We have presented a Bayesian calibration framework for the arterial Windkessel models. Our main interest is to obtain estimates of two physical interpretable parameters (total arterial compliance, $C$ and vascular resistance, $R$). For this reason, we have chosen as a modeling choice the WK2 (\ref{eq:WK2}). Acknowledging that this model cannot represent accurately high-frequency blood pressure waveforms we have accounted for systematic model discrepancy. 

To verify our modeling approach, through synthetic examples we have used a more complex model than the WK2 as the ground truth. First, we have considered 100 noisy  blood pressure simulations from the WK3 model for four different parameters setups in order to evaluate if the optimization approaches converge to the true value. We have obtained point estimates of the physical parameters by minimizing the residual sum of squares (RSS) between the observed data and the model predictions. Parameter estimates for the WK2 model for both $R$ and $C$ are biased and in most cases, the true values do not fall into the $90\%$ empirical confidence intervals. Interestingly, we found out that this holds in some cases for the WK3 model, where the data were simulated.

In a second experimental study, we used the WK3 model as ground truth with the same parameter setups as before, but now we considered a  case where we have three blood pressure cycles for each setup. We compared the optimization approaches with the Bayesian calibration using the WK2 as a modeling choice. The Bayesian calibration approach was able to correct the bias in the parameter estimates which is produced by using the WK2 model without accounting for model discrepancy. In most cases, it produced more accurate estimates than the WK3 model where data was simulated. This might reveal a possible identifiability problem for the WK3 model. Furthermore, the Bayesian calibration approach has recovered the true blood pressure waveform and reduced the uncertainty of subsequent model predictions.

It is important that Bayesian calibration allows for uncertainty quantification of the physical parameters though the posterior distributions. We found that in some cases parameter estimates can be quite uncertain with relatively wide $90\%$ credible intervals, especially for the total arterial compliance. This can make it harder to distinguish among hypertensives in some cases. Similar results were obtained for the real case study for two individuals, where the uncertainty in the posterior of the total arterial compliance was relatively large.

To obtain the posterior distributions of the physical parameters we used MCMC. This is computationally demanding since each MCMC evaluation requires the inversion of a large dense covariance matrix and this does not scale well in a digital twin setting. A remedy is to use a dimension reduction approach as proposed by e.g. \cite{higdon2008computer} which can reduce significantly the size of the covariance matrix. Gramacy et al. \cite{gramacy2015calibrating} proposed an attractive alternative by emulating the computer model using sparse local approximate GP regression \cite{gramacy2015local} and solving the calibration problem in the same formulation as proposed by \cite{kennedy2001bayesian} using derivative-free optimization while parametric bootstrap can be used to quantify the uncertainty \cite{kleijnen2014simulation}.

\section*{Acknowledgements}
We thank Kjell-Arne Øyen, Hans Olav Nilsen and Håvard Dalen for data collection in the real case study. We also want to thank all members of the My Medical Twin \href{https://www.ntnu.edu/digital-transformation/twin}{(MyMDT)}  project,  a digital twin for essential hypertension management and treatment.


\medskip
\bibliography{main}
	
\end{document}